\documentclass[structabstract]{aa}  
\usepackage{natbib}
\bibpunct{(}{)}{;}{a}{}{,} % to follow the A&A style 
\usepackage{graphicx}
\usepackage{txfonts}
\usepackage{amsfonts}
\begin{document}
\newcommand{\kmpsb}{km\,s$^{-1}$ }
\newcommand{\cdhob}{C$^{18}$O }
\newcommand{\cdsob}{C$^{17}$O }
\newcommand{\dzcob}{$^{12}$CO }
\newcommand{\tzcob}{$^{13}$CO }
\newcommand{\ndhpb}{N$_2$H$^+$ }
\newcommand{\cdb}{column density }
\newcommand{\ccb}{cm$^{-3}$ }
\newcommand{\cdeb}{cm$^{-2}$ }
\newcommand{\scb}{cm$^{-2}$ }
\newcommand{\ctdsb}{C$^{32}$S }
\newcommand{\ctqsb}{C$^{34}$S }
\newcommand{\tdsob}{$^{32}$SO }
\newcommand{\tqsob}{$^{34}$SO }
\newcommand{\juzb}{(J:1--0) }
\newcommand{\jdub}{(J:2--1) }
\newcommand{\jtdb}{(J:3--2) }
\newcommand{\denb}{n(H$_2$) }
\newcommand{\kmps}{km\,s$^{-1}$}
\newcommand{\cdho}{C$^{18}$O}
\newcommand{\cdso}{C$^{17}$O}
\newcommand{\dzco}{$^{12}$CO}
\newcommand{\tzco}{$^{13}$CO}
\newcommand{\ndhp}{N$_2$H$^+$}
\newcommand{\nddp}{N$_2$D$^+$}
\newcommand{\nddpb}{N$_2$D$^+$ }
\newcommand{\cd}{column density}
\newcommand{\cc}{cm$^{-3}$}
\newcommand{\cde}{cm$^{-2}$}
\newcommand{\ctds}{C$^{32}$S}
\newcommand{\ctqs}{C$^{34}$S}
\newcommand{\tdso}{$^{32}$SO}
\newcommand{\tqso}{$^{34}$SO}
\newcommand{\juz}{(J:1--0)}
\newcommand{\jdu}{(J:2--1)}
\newcommand{\jtd}{(J:3--2)}
\newcommand{\den}{n(H$_2$)}
\newcommand{\mjy}{MJy/sr}
\newcommand{\mjyb}{MJy/sr }
\newcommand{\Av}{A$_{\mathrm V}$}
\newcommand{\Avb}{A$_{\mathrm V}$ }
\newcommand{\SM}{M$_\odot$}
\newcommand{\SMb}{M$_\odot$ }
\newcommand{\pdix}[1]{$\times$ 10$^{#1}$}
\newcommand{\pdixb}[1]{$\times$ 10$^{#1}$ }
   \title{L1506\,: a prestellar core in the making
\thanks{based on observations made with the IRAM-30m.  IRAM is
   supported by INSU/CNRS (France), MPG (Germany), and IGN (Spain). }}
%   \subtitle{I. Overviewing the $\kappa$-mechanism}

   \author{ L. Pagani        \inst{1}
          \and
	 I. Ristorcelli \inst{2}
	 \and N. Boudet \inst{2}
	 \and M. Giard  \inst{2}
	\and A. Abergel \inst{3}
	\and J.-P. Bernard \inst{2}
	  }

\institute{LERMA \& UMR 8112 du CNRS,
 Observatoire de Paris,
 61 Av. de l'Observatoire, 75014 Paris, France\\
\email{laurent.pagani@obspm.fr}
 \and
 CESR \& UMR 5187 du CNRS/Universit\'e de Toulouse, 9 Av. du Colonel Roche, BP 4346, 31028 Toulouse Cedex 4, France\\
\email{ristorcelli@cesr.fr,giard@cesr.fr,bernard@cesr.fr}
 \and
  IAS,
   B{\^{a}}t. 121, Universit\'e Paris-Sud F-91435 Orsay\\
   \email{alain.abergel@ias.u-psud.fr}
           }

   \date{Received 09/07/2009; accepted 02/12/2009}

% \abstract{}{}{}{}{} 
% 5 {} token are mandatory
 
  \abstract
   {Exploring the structure and dynamics of cold starless clouds is necessary to understand the different steps leading to the formation of protostars. Because clouds evolve slowly, many of them must be studied \emph{in detail} to pick up different moments of a cloud's lifetime.}
   {
We study here a fragment of the long filament \object{L1506} in the Taurus region which we name \object{L1506C}, a core with interesting dust properties which have been evidenced with the PRONAOS balloon-borne telescope.}
% \emph{InfraRed Astronomical Satellite} (IRAS) colours.}
   {To trace the mass content of L1506C and its kinematics, we mapped the dust emission, and the line emission of two key species, \cdhob and \ndhp. \tzcob and \cdsob were also observed. We model the species emission using 1D Monte Carlo models.}
   {This cloud is reminiscent of L1498 but also shows peculiar features: i) a large envelope traced solely by \tzcob holding a much smaller core with a strong \cdhob depletion  in its center despite a low maximum opacity (\Av $\sim$ 20 mag), ii) extremely narrow \cdhob lines indicating a low, non-measurable turbulence, iii) contraction traced by \cdhob itself (plus rotation), iv) unexpectedly, the kinematical signature from the external envelope is opposite to the core one: the \tzcob and \cdhob velocity gradients have opposite directions and the \cdhob line profile is blue peaked on the contrary to the \tzcob one which is red peaked. The core is large (r = 3 \pdix{4} A.U.) and not very dense (n(H$_2$) $\leq$ 5 \pdix{4} \cc, possibly less). This core is therefore not prestellar yet.}
   {All these facts suggest that the core is kinematically detached from its envelope and in the process of forming a prestellar core. This is the first time that the dynamical formation of a prestellar core is witnessed. {The extremely low turbulence could be the reason for the strong depletion of this core despite its relatively low density and opacity in contrast with undepleted cores such as L1521E which shows a turbulence at least 4 times as high}.}


   \maketitle
%
%________________________________________________________________

\section{Introduction}

On their way to form stars, clouds go through a contraction phase leading to the formation of prestellar cores \citep[n $\geq$ 1 \pdix{5} \cc,][]{2008ApJ...683..238K} which are usually identified by their lack of internal heating sources, their large CO depletion, and their strong NH$_3$ and \ndhpb emission lines for which turbulence is often subthermal. These cores subsequently collapse to form protostars. Many details pertaining to these two condensation steps are still obscure and we don't understand what conditions are necessary to produce stars of different masses, high or low, what conditions are necessary to start the contraction of the cloud or the collapse of the prestellar core, nor the r\^ole of the magnetic field in supporting the clouds against collapse, etc... Obviously as the evolution time is long, one has to observe many clouds as a substitute to tracking any single cloud in the process of forming a star. For each cloud, physical and chemical properties should be studied in detail to assess all important parameters such as temperature, density, kinematics, and their respective gradients, if any, and also chemical abundances by volume.
%A combination of surveys and individual case studies is necessary. Surveys are useful to pick out interesting cases which deserve detailed studies like \object{L1521E} \citep{2004A&A...414L..53T}, or the famous \object{L1544} and \object{L183}. However, surveys must be used cautiously because the mass treatment of the data is often approximate and sometimes large errors in the derived parameters do occur. Therefore, a careful study of individual objects is a necessary condition to investigate the properties of dark clouds. This type of study should not be applied only to the most unusual or famous ones but to as many dark clouds as possible to sample their evolution towards forming new stars. 
By accumulating this type of studies, we can hope to disentangle depletion phenomenon versus cloud age or density, contraction versus oscillation, etc. and describe how prestellar cores and protostars form.

Many prestellar cores have been identified and some have already been studied in detail, including \object{B68}, \object{L183},  \object{L1498}, \object{L1544}, etc. Among these objects a few cases have been identified as \emph{early prestellar cores} if their density is already greater than 1 \pdix{5} \cc, or \emph{simple cores} probably about to form prestellar cores if not. The first type includes \object{L1495B}, \object{L1521B}, and \object{L1521E} \citep{2002ApJ...565..359H,2004ApJ...617..399H,2004A&A...414L..53T} which, despite a high density, show no sign of depletion (strong CO lines, weak or absent NH$_3$ and \ndhpb lines) and are considered as being both physically and chemically young while the second type is represented by two cases presently, \object{L1498} \citep{2004A&A...416..191T, 2005ApJ...632..982S} and {the very dense globule ' \object{G2}' in the Coalsack complex} \citep[][]{2004ApJ...610..303L}. L1498 in the study by  \citet{2005ApJ...632..982S} exhibits a low density (1--3 \pdix{4} \cc) and molecular depletion \citep[already reported by several other authors, especially][ who, by studying L1498,  reported the first \cdhob depletion in a prestellar core by comparison with a dust map]{1998ApJ...507L.171W}. The G2 {gas} content has not been studied in detail yet and only its peculiar, ring-like dust shape indicates its unstable nature despite its low density. This core is expected to turn into a prestellar core  \citep{2004ApJ...610..303L}. In this second category, which is considered as physically young but chemically evolved, we present a newcomer, a fraction of the L1506 filament.

The cloud we present in this paper, \object{L1506}, has attracted attention during a large-scale \tzcob survey \citep{1994ApJ...423L..59A} because of its low (though not unique) DIRBE\footnote{\emph{Diffuse Infra Red Background Experiment}, http://lambda.gsfc.nasa.gov} I$_{140\mu m}$/I$_{240\mu m}$  and IRAS\footnote{\emph{Infra Red Astronomical Satellite}, http://lambda.gsfc.nasa.gov} I$_{60\mu m}$/I$_{100\mu m}$ colour ratios to which the \tzcob maps were compared. Its dust properties were studied later on using the PRONAOS\footnote{\emph{PROgramme NAtional d'Observations Submillim\'etriques}} balloon-borne experiment continuum emission data in the submillimeter range (200 -- 600 $\mu$m) \citep{2003A&A...398..551S}, albeit with a limited angular resolution (2--3.5\arcmin). A significant change of the dust properties has been revealed going from the diffuse to the dense part of the cloud : no emission from transiently heated small particles, and a strong enhancement (by a factor of 3.4) of the submillimeter emissivity in the latter. This was interpreted as the signature of dust coagulation leading to the formation of fluffy aggregates (made of a mixture of very small and big grains). In order to understand this evolution and investigate what are the physical conditions associated with efficient aggregate formation (ice mantles, level of turbulence, densities,É), we are now concentrating on the study of the gas properties. It is presently clear that depletion plays a key r\^ole in the gas phase of dark clouds and basically two species are needed to map the clouds, namely \cdhob and \ndhp. NH$_3$ is also a good tracer but \ndhpb has some advantages over NH$_3$ among which the possibility to use the same telescope as for \cdho, giving a similar resolution and beam correction and also superior diagnostic capabilities as discussed by \citet{2007A&A...467..179P}. By observing these species, both the undepleted (outer) and the depleted (inner) cores can be traced and their kinematics and physical properties can be modeled. With \tzcob, we will also be able to trace a fraction of the extended envelope. We present and model such observations in this paper and discuss the properties of this cloud.

\section{Observations}

All observations were performed using the IRAM-30m telescope\footnote{http://www.iram-institute.org/}. The dark nebula L1506 is an elongated object \citep{1996ApJ...465..815O,1988A&A...189..207N} and the SIMBAD\footnote{http://simbad.u-strasbg.fr/sim-fid.pl} reference position ($\alpha_{2000}$ = 4$^h$18$^m$31.1$^s$ $\delta_{2000}$ = +25\degr 19\arcmin 25\arcsec) is outside the part of the cloud we present here which is centered on the PRONAOS emission peak \citep[$\alpha_{2000}$ = 4$^h$18$^m$50$^s$ $\delta_{2000}$ = +25\degr 19\arcmin 15\arcsec,][]{2003A&A...398..551S} which we propose to name \object{L1506C} \citep[A and B are already defined,][]{1999ApJS..123..233L}. Its distance is estimated to be 140 pc \citep{1978ApJ...224..857E,1994AJ....108.1872K}.

\begin{figure}
   \centering
\includegraphics[width=8cm,angle=-90]{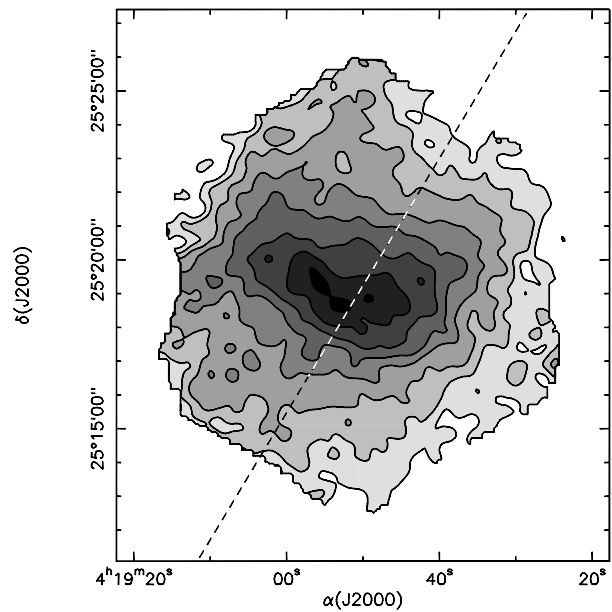}
   \caption{Dust emission at 1.2 mm as measured by MAMBO II. Resolution has been degraded to 30\arcsec\ for a better
   signal-to-noise ratio. 
   %Flux scale is in MJy\,sr$^{-1}$, levels 
   Levels are spaced by 0.5 MJy\,sr$^{-1}$ from 0.5 to 4 MJy\,sr$^{-1}$. The dashed line represents the cut observed with PRONAOS, and the white section the fraction of the cut observed with the IRAM-30m. The original cut direction was chosen to be perpendicular to the FIR filament (colour image on-line). The (0,0) position corresponds to $\alpha_{2000}$ = 4$^h$18$^m$50$^s$ $\delta_{2000}$ = +25\degr 19\arcmin 15\arcsec.
}
 \label{fig:mambo}%
  \end{figure}

In september 2004, we mapped the filament dust emission at 1.2 mm with the MAMBO\footnote{\emph{MAx-Planck-Millimeter-BOlometer, see IRAM website}} II bolometer. Skydips, pointings and calibration source observations were regularly performed, and data reduction was executed using IRAM proprietary software. The cold filament studied by \citet{2003A&A...398..551S} is clearly visible at the centre (Fig. \ref{fig:mambo}).
Spectroscopic observations were performed over several runs: 29 \& 30 November 2003, 22 \& 23 May 2004, 28 July 2004 and 18 July 2008. In all cases, the VESPA\footnote{\emph{VErsatile Spectroscopic  and Polarimetric Analyzer, see IRAM website}} autocorrelator was used with frequency sampling varying from 6 to 20 kHz for 3 mm lines and 40 kHz for 1.3 mm lines (Table \ref{tab:setting}). All observations were done in frequency switching mode. Receivers A\&B were used except in July 2008 where the 1 mm 9 pixel cameras HERA\footnote{\emph{HEterodyne Receiver Array, see IRAM website}} 1\&2 were used in the On-the-Fly mode, still using frequency switching. The \cdhob and \tzcob \juzb lines were observed simultaneously by tuning the receiver halfway between the two frequencies and offsetting the VESPA autocorrelator subwindows by $\pm$210 MHz in front of the two lines. Pointing was regularly checked and found to be stable within 3\arcsec. 

\begin{table}[htdp]
\caption{Telescope settings}
\label{tab:setting}
\centering
\begin{tabular}{ccccc}
\hline
Line&Frequency&Beam size& \multicolumn{2}{c}{Sampling}\\
&(MHz)&(\arcsec)&(kHz)&m\,s$^{-1}$\\
\hline
\nddpb\juz&77\,109.616$^\mathrm{a}$&32&6&23\\
\ndhpb\juz&93\,173.764$^\mathrm{a}$&27&10&32\\
\cdhob\juz&109\,782.176&23&10&27\\
\tzcob\juz&110\,201.360&23&20&54\\
\cdsob\juz&112\,358.982$^{\mathrm{b}}$&23&10&27\\
\cdhob\jdu&219\,560.358&12&40&54\\
\hline
\end{tabular}
\begin{list}{}{}
\item[$^{\mathrm{a}}$] New frequency for the main J$_\mathrm{FF\arcmin}$:1$_\mathrm{23} \rightarrow$ 0$_\mathrm{12}$ hyperfine component from \citet{2008arXiv0811.3289P}
\item[$^{\mathrm{b}}$] J$_\mathrm{F}$:1$_{7/2}$--0$_{5/2}$ transition from \citet{2003ApJ...582..262K}
\end{list}
\end{table}%

\begin{figure*}  
 \centering
\includegraphics[width=6cm,angle=-90]{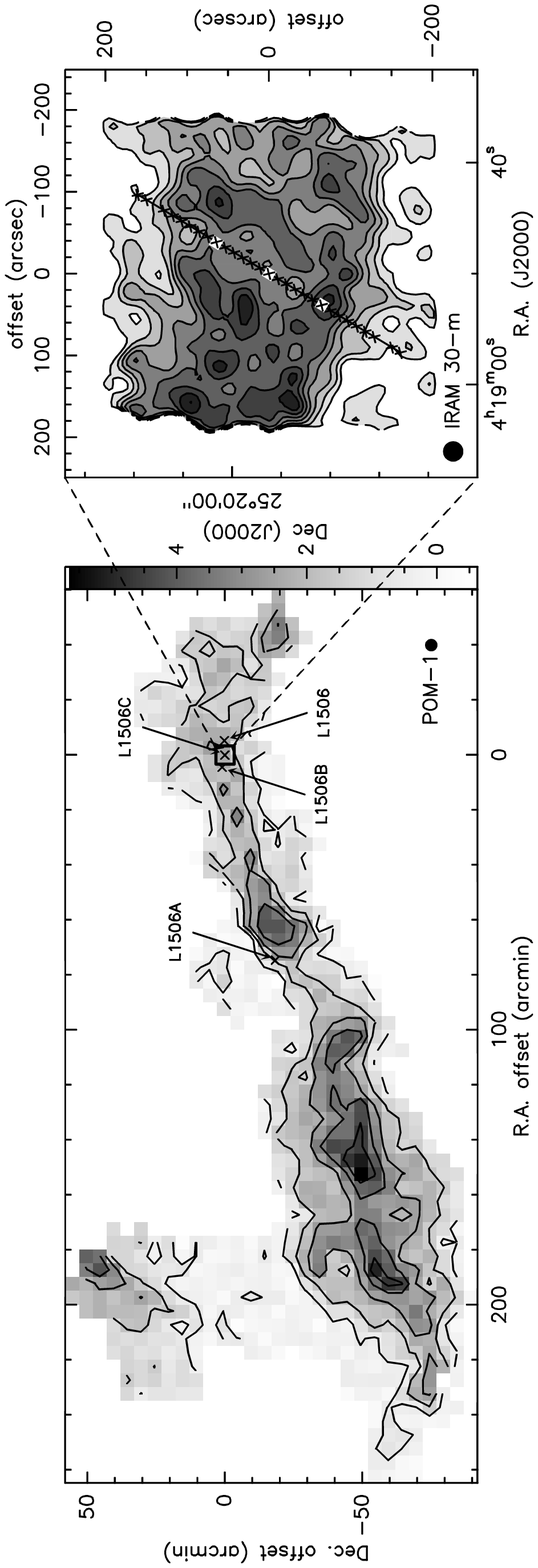}
   \caption{ {\bf Left}: low resolution \tzcob \juzb integrated intensity map of the L1506 Taurus cloud filament \citep[original data from][]{1988A&A...189..207N}. Contour levels are 0.5 to 4.5 K\,\kmpsb by steps of 1 K\,\kmps. The POM-1 2.5-m dish has a 4.5\arcmin\ resolution as shown in the bottom right corner. The square indicates the surface mapped in \cdhob \jdu. The offsets are with respect to L1506C coordinates (see text). {\bf Right}: the smoothed \cdhob \jdub integrated intensity map. Superposed to the map is the strip path observed with PRONAOS \citep{2003A&A...398..551S} along which we observed \tzco, \cdho, and \ndhpb with a good signal to noise ratio (see Fig. \ref{fig:coupe}). The crosses indicate the observed points and the 3 white boxes (superposed to 3 crosses) the points observed also in \cdsob (see Fig. \ref{fig:3pts}). The degraded 30-m beam size is represented. Levels are from 0.1 to 0.7 K\,\kmps\ by steps of 0.1 K\,\kmps. The (0,0) position corresponds to $\alpha_{2000}$ = 4$^h$18$^m$50$^s$ $\delta_{2000}$ = +25\degr 19\arcmin 15\arcsec.} 
   \label{fig:carte}%
  \end{figure*}
   \begin{figure*}[t]
   \flushleft
 \begin{minipage}[t]{12cm}
\includegraphics[width=12cm,angle=-0]{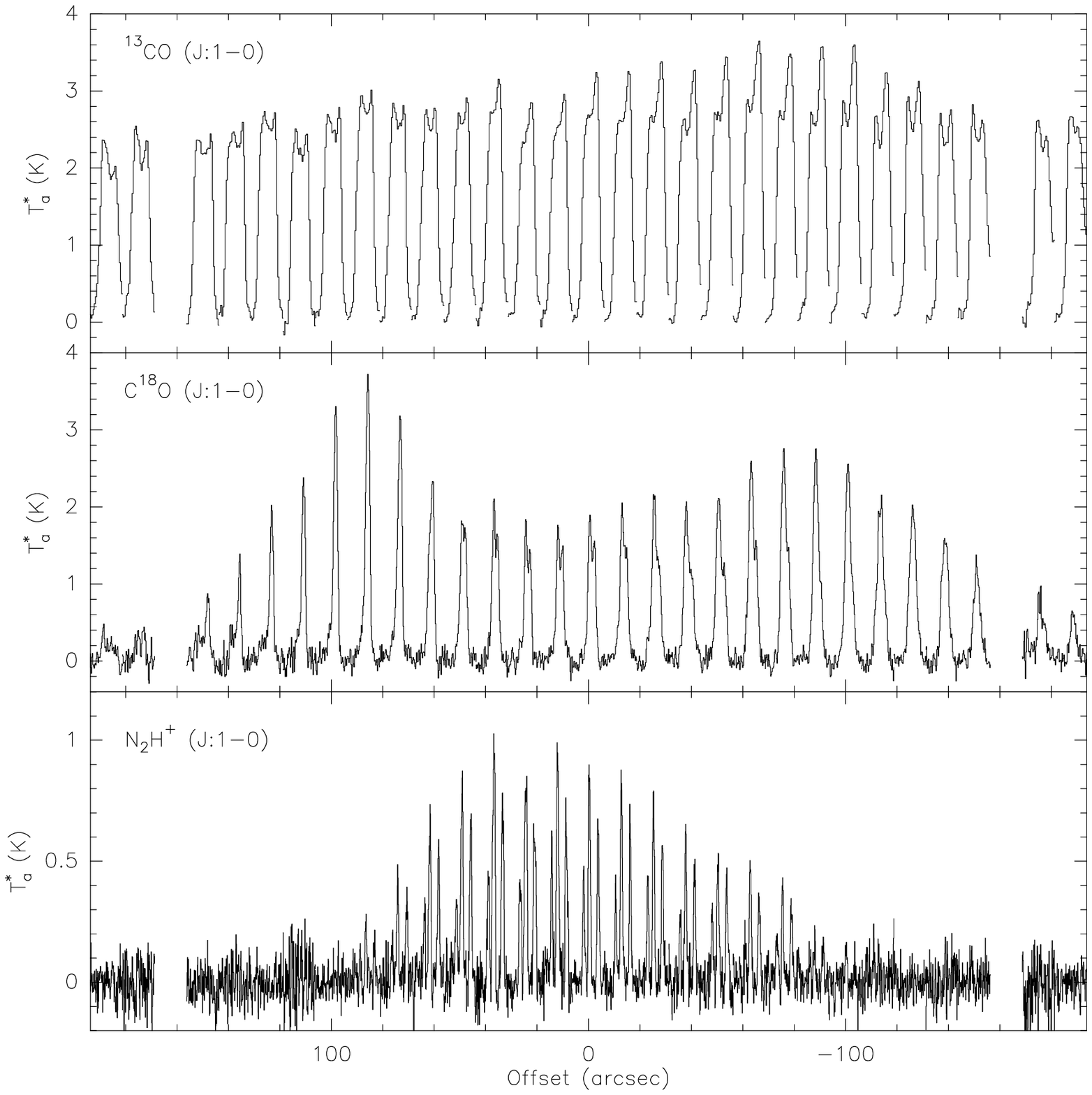}
\end{minipage}
  \hspace{0.5cm}
  \begin{minipage}[t]{5cm}
  \vspace{-5.5cm}
   \caption{Spectra taken along the PRONAOS cut (see Fig. \ref{fig:carte}). The upper row shows the \tzcob \juzb spectra. The middle row shows the \cdhob  \juzb spectra. The lower row shows the \ndhpb  \juzb spectra. The velocity scale is 6.5 to 8 \kmps\ for \tzcob and \cdho, and 5.5 to 9 \kmps\ for \ndhp. Offsets increase from North-West to South-East.}
\label{fig:coupe}% 
\end{minipage}
\end{figure*}
 
\section{Results and analysis}

\begin{figure}[t]
\centering
\includegraphics[width=6cm,angle=-90]{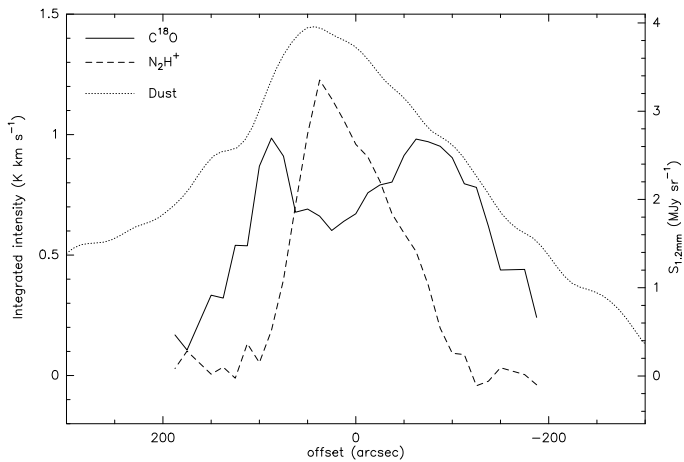}
\caption{Integrated intensity of \cdhob and \ndhpb \juzb compared to the dust emission as measured by MAMBO along the PRONAOS cut. MAMBO data are smoothed to 40\arcsec\ here. The core is clearly not symmetrical and \cdhob depletion is obvious.}
\label{fig:coupedust}
\end{figure}

\begin{figure}[tbp]
\centering
\includegraphics[width=8cm]{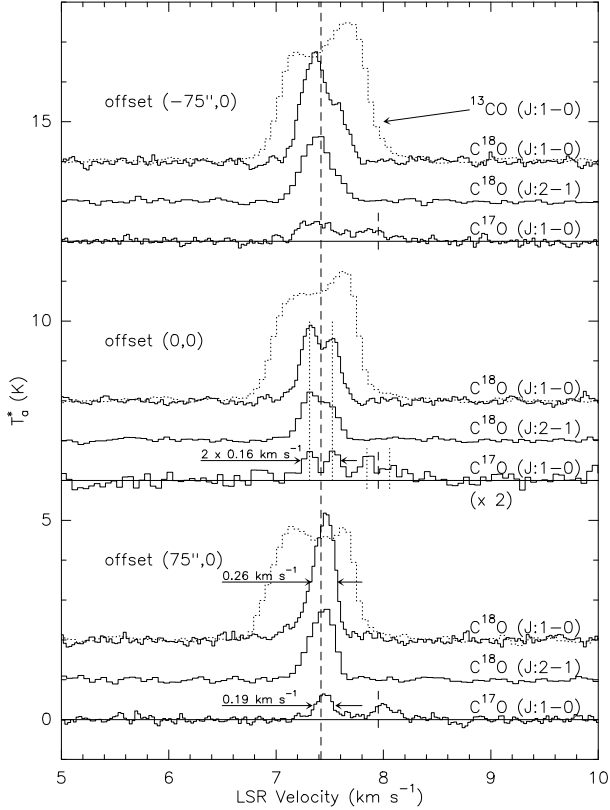}
\caption{Spectra of \tzcob (dotted lines), C$^{17}$O, and C$^{18}$O  taken at 3 different positions along the PRONAOS strip: in the middle and close to each side peak \cdhob emission. The full width at half maximum is indicated for some remarquably narrow lines. Only the \cdsob  J$_\mathrm{F}$:1$_{7/2}$--0$_{5/2}$ and  J$_\mathrm{F}$:1$_{3/2}$--0$_{5/2}$ are displayed. The dashed line marks the systemic velocity (with an offset of +0.534 \kmpsb for the J$_\mathrm{F}$:1$_{3/2}$--0$_{5/2}$ \cdsob transition) while the dotted lines for the \cdsob transition show the symmetrical velocity displacement of each \cdsob hyperfine component in the (0,0) direction (see text). Lines are artificially offset vertically and the central offset \cdsob line is multiplied by a factor of 2. The \cdhob \jdub lines are shown at their original 11\arcsec\ resolution.}
\label{fig:3pts}
\end{figure}

Figure \ref{fig:carte} shows a large-scale \tzcob \juzb map obtained with  POM-1\footnote{\emph{Petite Op\'eration Millim\'etrique}} \citep[a 2.5 m dish telescope  with 4.5\arcmin\ resolution and 100 kHz filters -- 0.27 \kmpsb --  which was in service at Bordeaux Observatory. Original data taken from][]{1988A&A...189..207N}, and the region of interest is delineated by a square. This map traces the large-scale envelope of the cloud. On the right handside,  the \cdhob \jdub integrated intensity map obtained on-the-fly with the 30-m over L1506C is shown. We degraded the resolution to 24\arcsec\ to improve the signal-to-noise ratio. A large hole in emission is visible in the centre of the map. This is due to CO depletion, as is now well established. The PRONAOS strip centered at the 100 $\mu$m emission peak {came across} the most depleted part of the filament.
This depletion is also spectacular in terms of spectrum peak temperature (Fig. \ref{fig:coupe}) or integrated intensity (Fig. \ref{fig:coupedust}) along the PRONAOS strip. In these figures, the drop of \cdhob intensity is correlated with the appearance of \ndhp, another sure sign of CO depletion as CO readily destroys \ndhpb \citep{1997ApJ...486..316B, 2002P&SS...50.1133C, 2005A&A...429..181P}. To check that the \cdhob lines are not dropping for opacity reasons in the centre of the cloud, we observed 3 positions in \cdsob \juzb. We chose two positions close to the \cdhob peaks on each side of the strip ($\pm$75\arcsec) and one in the middle, close to the local minimum of the \cdhob emission (Fig. \ref{fig:3pts}). These 3 positions are indicated on Fig.  \ref{fig:carte} by 3 small white boxes superposed to 3 of the crosses along the PRONAOS strip. Several interesting features are visible: 

\begin{enumerate}
\item the (0, 0) and (+75\arcsec, 0) offset positions (along the cut) display extremely narrow \cdsob and \cdhob lines (FWHM = 0.16 to 0.26 \kmpsb where FWHM stands for full width half maximum). 
\item the existence of a velocity gradient across the strip which possibly traces rotation in the \cdhob data.
\item the existence of another velocity gradient for \tzcob of \emph{opposite} direction.
\item the \cdhob \juzb splitting towards the reference position, which is not so much visible in the \jdub line but is clearly so in the \cdsob \juzb line, as the dotted lines reveal in Fig. \ref{fig:3pts}.
\end{enumerate}

As most of the properties of this cloud change gradually with radius, it is difficult to define what is the core and what is the envelope. One parameter however changes abruptly and this is the velocity field. We will therefore define the envelope as the region where \tzcob is revealing a different velocity field from the other species while \cdhob and \ndhpb both trace the core, the outer, undepleted part for the former and the inner depleted one for the latter.

As far as we know, these CO narrow lines are the narrowest C$^{18}$O and \cdsob lines reported so far, comparable to the very narrow features seen in NH$_3$ and \ndhpb in sources like L183 \citep{2007A&A...467..179P}. The main difference is that these \cdhob and \cdsob lines are emitted outside the depleted core, in a region where turbulence is usually still high. The lines are narrow enough so that the \cdsob J$_\mathrm{F}$:1$_{7/2}$--0$_{5/2}$ and  J$_\mathrm{F}$:1$_{3/2}$--0$_{5/2}$ hyperfine components which are usually blended, are completely separated here. 

\begin{figure}[tbp]
\centering
\includegraphics[width=6.5cm,angle=-90]{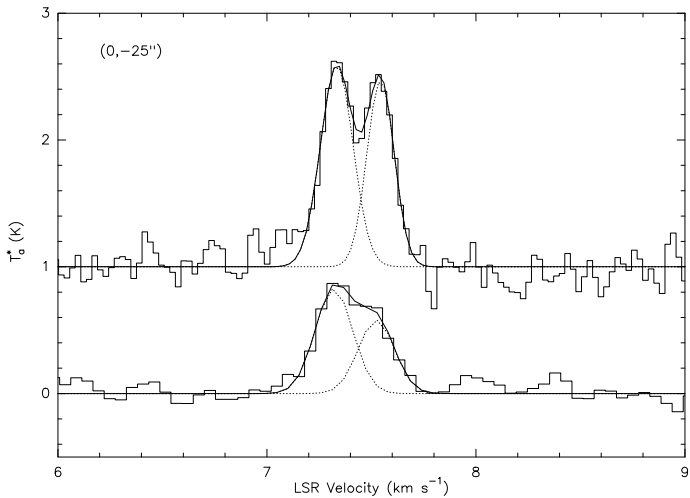}
\caption{\cdhob \juzb (top) and \jdub (bottom) spectra (histogram plot) taken 25\arcsec\ away from the PRONAOS cut. The total Gaussian fit (full line) for both transitions is shown as well as individual components. The components are separated by 200 m\,s$^{-1}$ and their width is 150 (right) and 180 (left) m\,s$^{-1}$}
\label{fig:1pt}
\end{figure}

{Figure \ref{fig:1pt} confirms the extraordinary narrowness of the lines towards the centre of the cloud. These \cdhob spectra have been taken at (0,-25\arcsec) in the frame of the PRONAOS cut, that is (-21.6\arcsec, -12.6\arcsec) in the Equatorial J2000 frame. The \juzb spectrum shows clearly two components which we have fitted with 2 independent gaussians. We have also fitted the \jdub spectrum with two unconstrained gaussians. The velocity of each of the two components agree for both lines within the uncertainty of the fit (a difference of 10 m\,s$^{-1}$ for an uncertainty of 30 and 40 m\,s$^{-1}$ for the left and right components, respectively).
As for \cdsob in Fig. \ref{fig:3pts}, the splitting between the two components is $\sim$200 m\,s$^{-1}$ and the width of the individual components is 150 (right) and 180 (left) m\,s$^{-1}$. For a standard gas kinetic temperature of 10 K, {for which the optically thin thermal linewidth of CO is $\sim$0.12 \kmps,} this sets the turbulent width contribution to the linewidth to $\lesssim$90~m\,s$^{-1}$.}

If the \cdhob and \cdsob splitting was due to self-absorption, it would be exactly the reverse, the \cdhob \jdub line would have had the deepest self-absorption feature and normally no such feature would have been seen in \cdsob since, due to its low abundance and to its hyperfine structure, it is very difficult to make it optically thick. 
%Because the beam is relatively small compared to the cloud, if these symmetrical features were due to rotation, it would be rotation very near the rotational axis since, far from the axis but still on the line of sight going through it, the rotational components picked up by the beam are perpendicular to the line of sight and therefore do not contribute in terms of Doppler shifting. 
These symmetric features cannot be due to rotational D\"oppler shifting because for a narrow beam aiming at the rotational axis, most of the material picked up by the beam is moving at right angle to the line of sight having therefore no contribution to the velocity component along it. Only material very near the rotational axis would contribute to the D\"oppler shifted emission.
As the inner core is depleted in CO, such a contribution near the axis is not possible and therefore rotation cannot explain this splitting. To explain this double peak feature, the only  possibility left is that the CO envelope is radially moving. The direction of movement (expansion or contraction) can be told from the asymmetry of the \cdhob \juzb and \jdub lines. Here, these \cdhob lines are stronger in the blue component than in the red component. {As we will confirm with the Monte-Carlo model (Sect. 4), this can only be explained by contraction and not expansion despite the absence of heavy self-absorption which is usually invoked to justify blue peaked profiles \citep{1996ApJ...465L.133M}. The lines being almost independent due to their narrowness, the blue and red line contributions are emitted from only one side of the cloud each -- front or back depending on the movement, expansion or contraction -- and have no or very little interaction.  The line which comes from the rear part is originating in the backside low density envelope and gets contribution from a material getting denser as the photons travel towards the observer, until they reach the depleted zone and then cross the rest of the cloud without interaction. For a constant kinetic temperature, this means that the excitation temperature of the contributing material is increasing (or remaining constant if thermalized) towards the observer and the successive layers add their contribution to the emergent signal. Conversely, the line formed on the near side of the cloud starts from the highest density and excitation temperature layer, just outside the depleted zone and gets contributions from lesser and lesser dense material with therefore a decreasing excitation temperature. When the excitation temperature decreases, net absorption of photons can happen if opacity is not null. Indeed, the \cdhob lines are not really optically thin (opacities are in the range of 0.1 to 0.5) and this is enough to slightly differentiate between the line which comes from the rear part of the cloud which has undergone no attenuation after the highest excitation temperature layer emission and the line which comes from the front part and has suffered from a small absorption due to the outer, less excited layers. {The effect is clearly seen towards the (0,0) position : the \cdhob \juzb line is 20\% stronger on the blue side (1.81 and 1.48 K peak temperature for the blue and red components, a difference greater than 10 $\sigma$, with  $\sigma$ = 27 mK), and the \cdhob \jdub line which due to higher opacity and higher sensitivity to density conditions has lost its red peak (Fig. \ref{fig:3pts})}. The red component is therefore the one emitted from the front part of the cloud, the blue component from the rear part and this indicates that the cloud is contracting rather than expanding\footnote{There exists the alternative possibility that the blue peak is stronger than the red peak because of a difference in column densities between front and rear parts of the cloud. If it were the case in such a manner that it would hide the cloud expansion, the difference in column density between the two components would be $\sim$40\% in the optically thin limit (and even more with growing opacity). Indeed, as the intensity difference is about 20\% towards the (0,0) position in \cdhob \juz, it would have to be compensated twice to go from a stronger red component (due to expansion combined with differential excitation) to a weaker red component (as observed) and the effect would be more pronounced for the \jdub component. This possibility seems therefore quite unlikely and even if the two sides of the cloud have different column densities, contraction remains the best explanation.}. }
The amount of contraction of the \cdhob core is indicated by the two dotted lines in Fig. \ref{fig:3pts}, namely $\approx \pm$100 m\,s$^{-1}$. Therefore, the \cdhob outer core both rotates {\bf(from the ($\pm$75\arcsec,0) symmetrical velocity offsets seen in Fig. \ref{fig:3pts} as mentioned above}) and contracts. To our knowledge, this is the first time that core contraction is observed based on \cdhob and \cdsob lines.

\begin{figure*}[htbp]
\centering
\includegraphics[width=6.8cm,angle=-90]{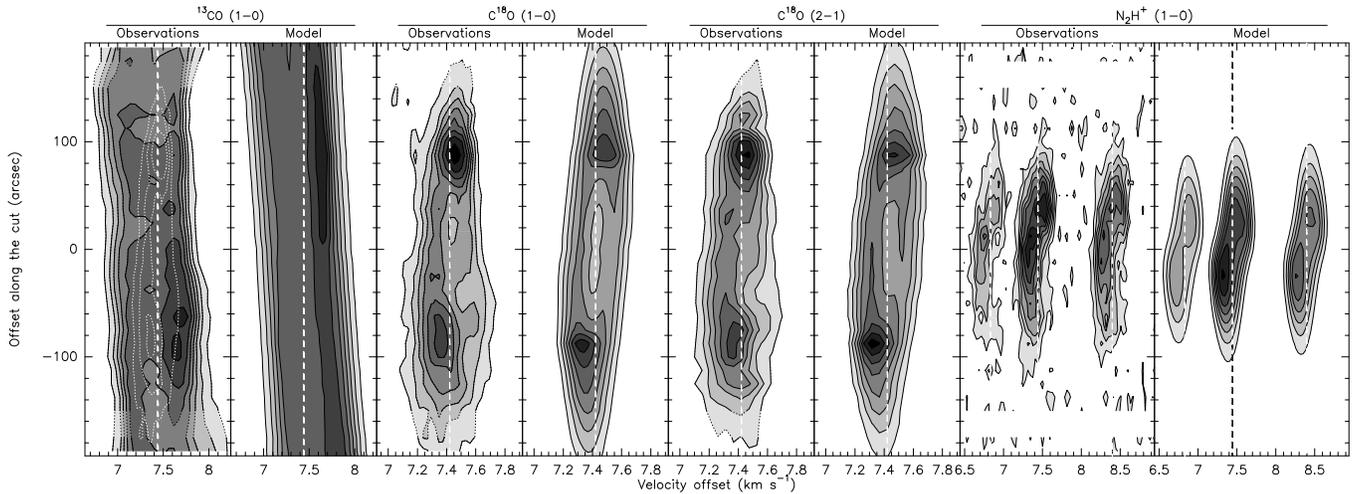}
\caption{Position-velocity plots along the PRONAOS cut. For each line, we show the observations on the left and the model on the right with the same contour levels. For \ndhp, we reproduce only the 3 inner hyperfine components. The dashed line marks the systemic velocity of the cloud. For \tzcob and \cdhob observations, the dotted contours indicate interpolated data (data is missing for offsets $\pm$162.5\arcsec). Superposed to the \tzcob observations, dotted contour levels from the \cdhob \juzb observation plot are repeated. Only levels 1, 2, and 3 K have been drawn for clarity.}
\label{fig:VPobsmod}
\end{figure*}

The \ndhpb hyperfine structure lines do not have such a good signal-to-noise ratio and do not show a structured profile. However, their displacement in velocity along the cut is similar to that of \cdhob as we can see in Fig. \ref{fig:VPobsmod}. 

The \tzcob \juzb line which was observed simultaneously with the \cdhob \juzb line shows a very different behaviour (Figs. \ref{fig:coupe}, \ref{fig:3pts}, \& \ref{fig:VPobsmod}). Its peak intensity is red-shifted and its velocity gradient along the cut is opposite to that of the \cdhob line.
From the large scale \tzcob map obtained with POM-1, we know that the L1506 filament has a width of $\sim$40\arcmin\ in the region around L1506C (Fig. \ref{fig:carte}). This is $\sim$6 times larger than the \cdhob extent. As we do not know how the \tzcob velocity behaves beyond these central 6\arcmin, we cannot claim that this velocity drift is the signature of a rotation of the large scale filament but it definitely indicates that it is opposite to the \cdhob rotation (Figs.  \ref{fig:3pts} \& \ref{fig:VPobsmod}). Similarly, the red peak in the \tzcob velocity profile is probably a sign of gas expansion (Figs. \ref{fig:coupe} \& \ref{fig:3pts}), while \cdhob is tracing a contraction. \emph{Therefore the extended envelope is dynamically behaving at the opposite of the core}.

Figure \ref{fig:coupedust} shows the dissymmetry of the cloud in both dust and gas. It is remarkable that the \cdhob lines on each side of the cut have the same maximum integrated intensity despite a variation of 25\% in peak intensity. It indicates that the quantity of gas is comparable and that the main reason for the intensity change is the line width difference. It must also be noticed that the \ndhpb integrated intensity profile is strongly asymmetric and varies abruptly at the top, two features we cannot reproduce with our standard one-dimensional (1D) model (see Sect. \ref{models}). {\ Its peak position is close to but not coincident with the dust peak position}.

We observed the  position with the {second} strongest \ndhpb integrated intensity (offset +25\arcsec\ along the PRONAOS strip) in \nddpb \juzb to check the level of deuteration in the inner core {(the strongest peak -- offset +37.5\arcsec\ -- was identified afterwards)}. We found only a very weak signal (Fig. \ref{fig:n2dp}), the main hyperfine component being only 70 mK. A hyperfine fit was performed using the CLASS fitting routine MINIMIZE with the HFS option\footnote{http://www.iram.fr/IRAMFR/GILDAS/}.

\section{Models}\label{models}

\subsection{density profile}

{The density radial distribution has been derived from the MAMBO map emission
using a Wiener linear inversion method as described by  \citet{2002MNRAS.330..497D}. 
{Considering the elongated distribution (along the RA axis), we have
modelled the filament with a cylindrical geometry approximation, with its axis in the plane of the sky and perpendicular to the PRONAOS strip, as in   \citet{2003A&A...398..551S}. }
We have also taken into account a temperature inward decrease from 15\,K down to 8\,K. This
range has been deduced from the Av profile in the filament  \citep{2003A&A...398..551S} combined
with the prediction from  \citet{2001A&A...376..650Z} or \citet{1992A&A...263..258B}  dust temperature models in cold cores. 
{We have discretized the cylinder into iso-density and iso-temperature rings, and determined
the array A such as: 
$$N_{H2}(j) = A(i,j) \times n_{H2}(i) + b$$ or 
$$I_{\lambda}^{j} = \sum_{i} A(i,j) \times B_{\lambda}(T_{i}) \times \kappa_{1200~\mu m} \times m_{H} \times \mu \times n_{H2}(i)  + b $$
where (i,j) are respectively the indexes for the rings and the lines of sight, N$_{H2}$ the column density,  n$_{H2}$(i) the gas density in the ring i,  $I_{\lambda}^{j} $ the observed intensity along the line of sight j, b is the noise level, B$_{\lambda}$(T$_{i}$) is the black-body function evaluated for the temperature T of the ring i, m$_H$ is the proton mass, $\mu=2.33$ is the mean molecular weight of interstellar material in molecular clouds and $\kappa_{1200~\mu m}=8 \times 10^{-3}  \mathrm{cm}^{-2}\,  \mathrm{g}^{-1}$ is the mass absorption coefficient assumed constant in the core and derived from the dust aggregates emissivity model by  \citet{1994A&A...291..943O}.  
The local density n$_{H2}$  in each ring (i) is deduced from : 
$n_{H2} = W \times I_{\lambda} $, where we adopt the optimized inversion matrix : $$W = {(^{t}A' \times A'})^{-1} \times  {^{t}A' }$$ 
(cf Dupac $\&$ Giard (2002)), with: \\
$$A'= \kappa_{1200~\mu m} \times m_{H} \times A(i,j) \times  B_{\lambda}(T_{i})$$ 
}
The best optimization has been obtained using a ring thickness value
of 50\arcsec, and the result leads to an inner density of 2.4 \pdix{4} \cc\  (in average over a 50\arcsec\  
radius core).

\subsection{line modeling}

{The density profile derived from the dust data has been used as a first guess for our 1D line radiative transfer models but we allowed the profile to be changed because the raw dust map is noisy and its final resolution for the inversion method is only 50\arcsec\ compared to the \cdhob \juzb 24\arcsec\ and \jdub 12\arcsec\ and because the dust properties are not strongly constrained and can vary, therefore making the derived density uncertain by a factor of 2 typically.}

For all three species, \tzco, \cdho, and \ndhp, we fit a non-local thermodynamic equilibrium radiative transfer Monte-Carlo model and try to reproduce the spectra (Figs. \ref{fig:fitc18o10}, \ref{fig:fitc18o21}, \ref{fig:fit13co10}, \& \ref{fig:fitn2hp10}) and the velocity-position diagram (Fig. \ref{fig:VPobsmod}). The 1D Monte-Carlo model was originally developed by \citet{1979A&A....73...67B}. It includes microscopic turbulence and a radial velocity field. It was subsequently modified to include rotation \citep{1996A&A...312..989P} and a variant was developed to treat the hyperfine structure of \ndhpb \citep{2007A&A...467..179P}. The 1D cloud parameters are shown in Table \ref{tab:MC} and the species abundances are traced as a function of radius in Fig. \ref{fig:profil}. {Here the model is spherical, with a radius equal to the dust cylinder radius and with the rotation axis similar to the cylinder axis of the dust model, i.e. supposed to be aligned with the embedding filament elongation and therefore perpendicular to the PRONAOS strip. The model data to be compared to the observations are taken at the equator considered to be a good approximation of a cylinder slice, as already discussed in \citet{2007A&A...467..179P} for a similar case}. For the \tzcob emission, as our high spatial and velocity resolution strip only extends upon 6\arcmin\ and because the \tzcob line is optically thick and much more extended, it is difficult to build a simple 1D model to reproduce and constrain the 30-m \tzcob observations (the POM-1 data are too coarse and too noisy for this purpose). More generally, the cloud is not symmetrical and therefore our 1D Monte-Carlo model cannot reproduce the various differences between positive and negative offsets. However, we have tried to fit globally the observations and overall to use a similar cloud description for the \ndhpb and \cdhob line models. Though the \tzcob data are too difficult to reproduce with the 1D Monte-Carlo code, we have attempted to fit them as well, at least to get some estimate of the physical conditions in the outer parts which could explain the observations. From the \cdhob and \ndhpb data, we find that the cloud rotational axis could also be slightly offset along the PRONAOS cut (about +12\arcsec) with respect to our reference position (which was set somewhat arbitrarily). The peak \ndhpb integrated intensity is even further away from the reference position (+37.5\arcsec) but setting the rotational axis that far does not fit with the overall velocity profile. Despite the fact that the cloud extension on the positive offset side is smaller than on the negative offset side, the fit is relatively good, especially towards the centre of the cloud. {As mentioned above,} we have tried to somewhat adjust the density profile and found a solution which is close to \citet{2002ApJ...569..815T} formula for the \ndhpb-- \cdhob region :

\begin{equation}
n(\rm{H_2}) = \frac{n_0(\rm{H_2})}{1 + \left(\frac{r}{r_0}\right)^\alpha}\hspace{0.2cm} cm^{-3}
\end{equation}

\begin{figure}[t]
\centering
\includegraphics[width=6.4cm,angle=-90]{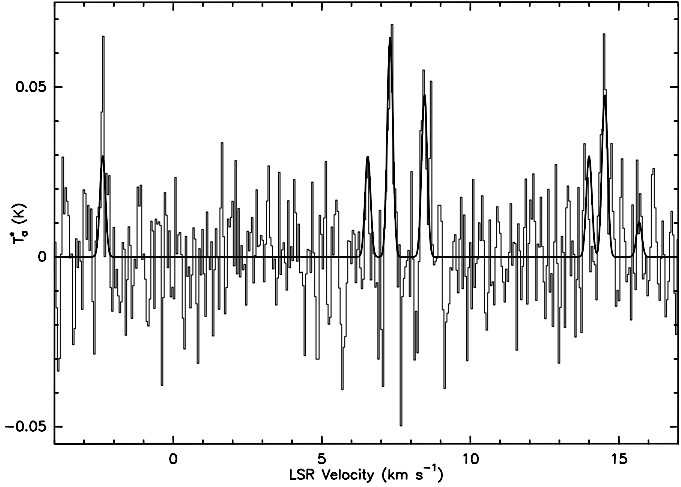}
\caption{Spectrum of \nddpb taken at offset +25\arcsec along the PRONAOS strip. The hyperfine fit is superposed to the data. The data is Hanning smoothed once. The sampling is 50 m\,s$^{-1}$ and the rms is 13 mK.}
\label{fig:n2dp}
\end{figure}

\noindent with n$_0$(H$_2$) = 5 \pdix{4} \cc, r$_0$ =  1 \pdix{4}A.U. and $\alpha$ = 2.5. This profile provides a peak dust emission comparable to the MAMBO observation, and the average density of $\sim$4 \pdix{4} \ccb in a 50\arcsec\ radius is within a factor of 2 of the peak density derived from the MAMBO observations. Apart from the dust properties which are not strongly constrained, an equivalently possible explanation for this factor of $\lesssim$ 2 difference is that the He + \ndhpb collisional coefficients we presently use \citep{2005MNRAS.363.1083D} are overestimating the true H$_2$ density needed to collisionally excite the \ndhpb to its J =  1 level as discussed in \citet{2007A&A...467..179P}. {From dust and \ndhpb measurements, we find that the density in this core is rather low, a factor of 2 to 4 below the threshold proposed by \citet{2008ApJ...683..238K} for prestellar cores. Therefore, it is not such a core.}

 \begin{figure*}[htbp]
   \flushleft
 \begin{minipage}[t]{12cm}
\includegraphics[width=8.5cm,angle=-90]{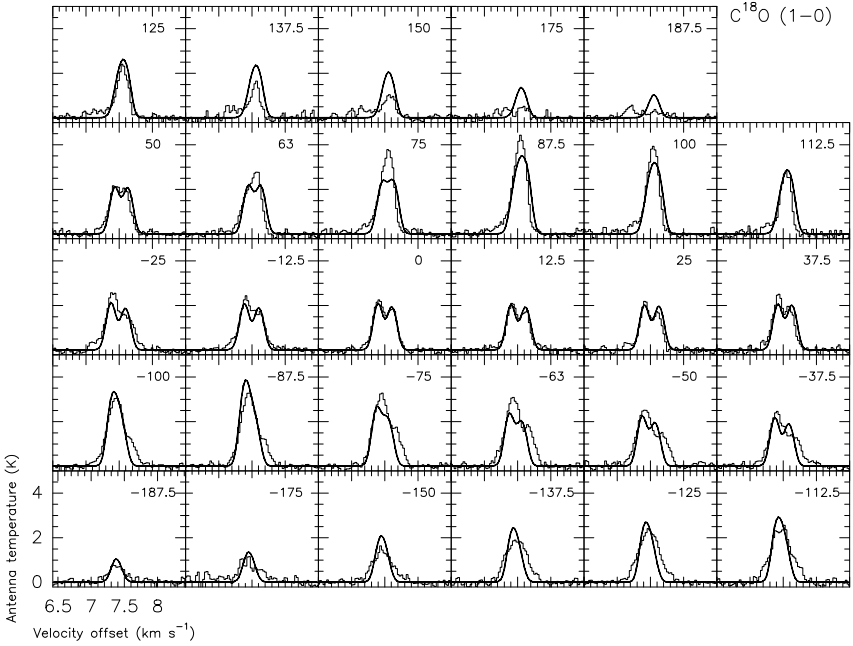}
\end{minipage}
  \hspace{0.5cm}
  \begin{minipage}[t]{4.5cm}
  \vspace{4.5cm}
\caption{Fit of the individual \cdhob \juzb spectra along the PRONAOS cut. The histogram plot represents the data, the thick continuous curve, the model. A FWHM turbulent velocity of 0.11 \kmpsb has been used all over the cut which explains why the model lines are larger than the observations for the positive offsets. Offsets along the PRONAOS cut in arcseconds are given in the upper right corner of each spectrum.}
\label{fig:fitc18o10}
  \end{minipage}
\end{figure*}

\begin{figure*}[htbp]
   \flushleft
 \begin{minipage}[t]{12cm}
\includegraphics[width=8.5cm,angle=-90]{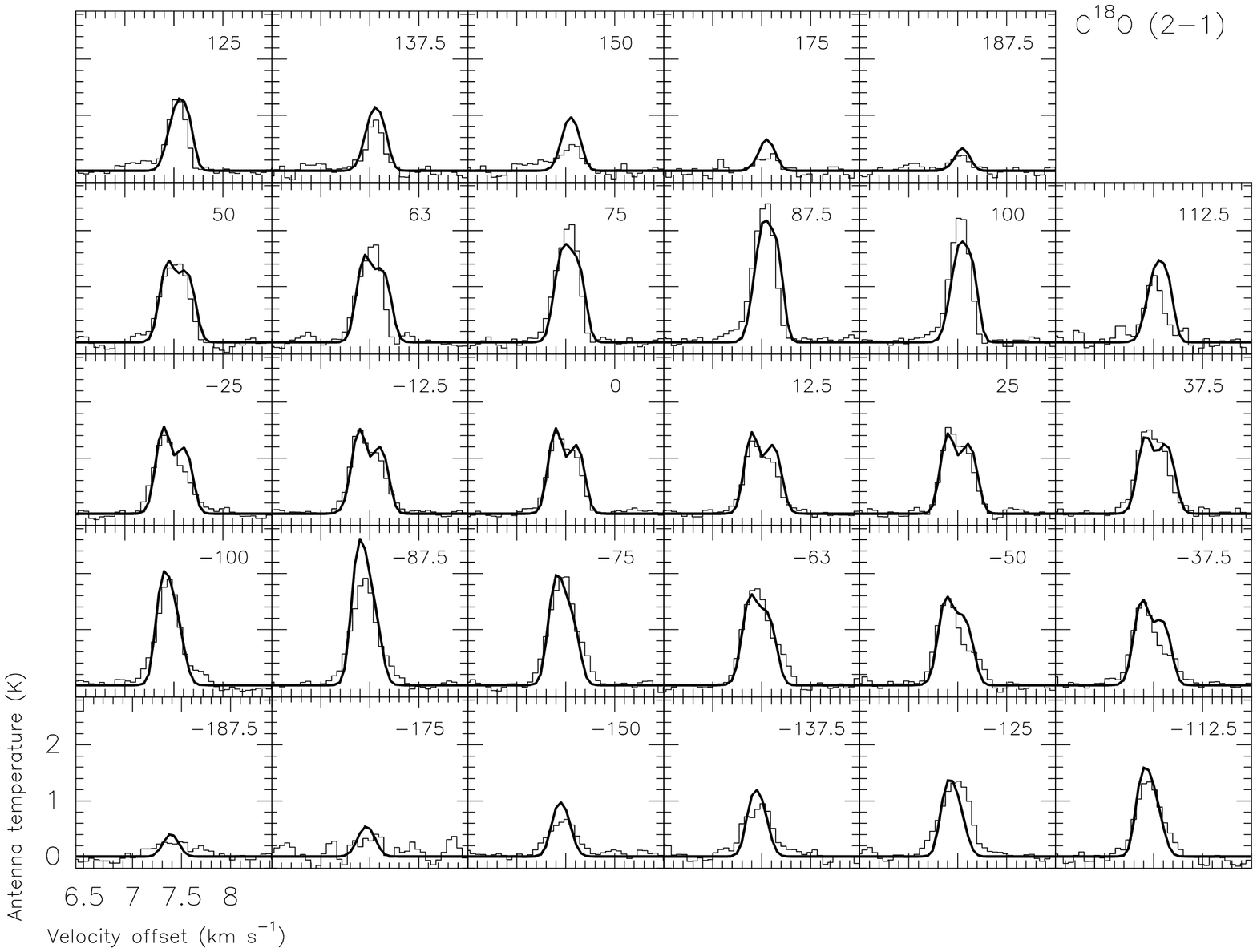}
\end{minipage}
  \hspace{0.5cm}
  \begin{minipage}[t]{4.5cm}
  \vspace{5.5cm}
\caption{Same as Fig. \ref{fig:fitc18o10} for \cdhob \jdub spectra. }
\label{fig:fitc18o21}
  \end{minipage}
\end{figure*}

\begin{figure*}[htbp]
   \flushleft
 \begin{minipage}[t]{12cm}
\includegraphics[width=8.5cm,angle=-90]{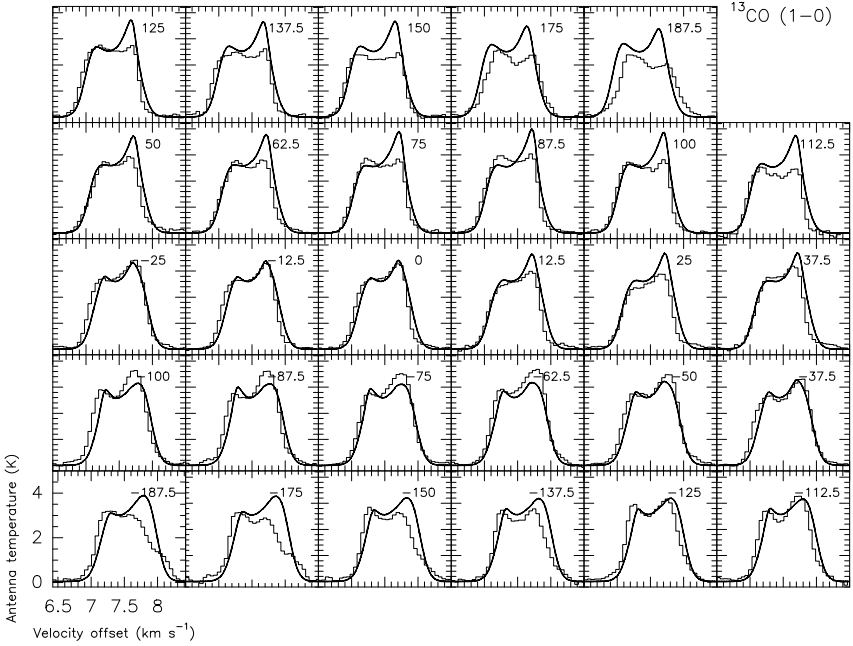}
\end{minipage}
  \hspace{0.5cm}
  \begin{minipage}[t]{4.5cm}
  \vspace{5.5cm}
\caption{Fit of the individual \tzcob \juzb spectra along the PRONAOS cut. The histogram plot represents the data, the thick continuous curve, the model.}
\label{fig:fit13co10}
  \end{minipage}
\end{figure*}

\begin{figure*}[htbp]
   \flushleft
 \begin{minipage}[t]{12cm}
\includegraphics[width=8.5cm,angle=-90]{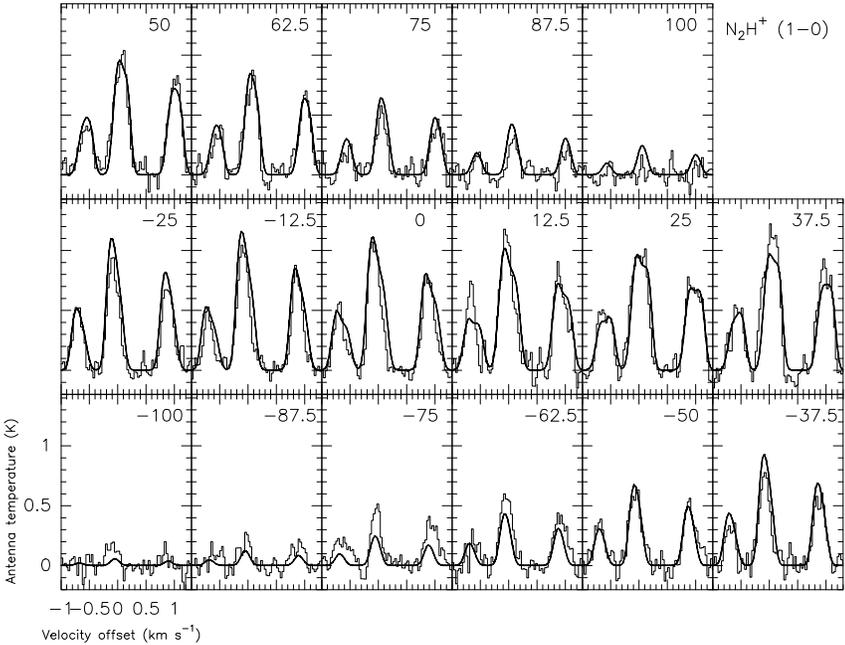}
\end{minipage}
  \hspace{0.5cm}
  \begin{minipage}[t]{4.5cm}
  \vspace{4.5cm}
\caption{Fit of the individual \ndhpb \juzb spectra along the PRONAOS cut. The histogram plot represents the data, the thick continuous curve, the model. Only the 3 central hyperfine structure components are displayed. A turbulent velocity of 0.068 \kmpsb has been used.}
\label{fig:fitn2hp10}
  \end{minipage}
\end{figure*}

\begin{table*}[htdp]
\caption{The Monte-Carlo cloud model used to fit the spectra (Figs. \ref{fig:VPobsmod}, and \ref{fig:fitc18o10} -- \ref{fig:fitn2hp10})}
\label{tab:MC}
\centering
\begin{tabular}{cccccccccc}
\hline
\multicolumn{2}{c}{Shell Radius}&Density&Temperature& Rad.velocity$^{\mathrm{a}}$&Rot.velocity&Turbulence$^{\mathrm{b}}$&\multicolumn{3}{c}{Abundance$^{\mathrm{c}}$}\\
\cline{1-2}\cline{8-10}
(A.U.)&(\arcsec)&(\cc)&(K)&(\kmps)&(\kmps)&(\kmps)&\ndhp&\cdho&\tzco\\
\hline
1740 & 12 &  50000  & 10 &   0.11 &  0.200 &  $\leq$0.11 / $\leq$0.068 &  1.5(-11) &  $\leq$5(-9)&$\leq$3(-8)\\
3480 & 25 &  47400  & 10 &   0.11 &  0.141 &  $\leq$0.11 / $\leq$0.068 &  2.5(-10) & $\leq$ 5(-9)&$\leq$3(-8)\\
5220 & 37 &  41300  & 10 &   0.11 &  0.115 &  $\leq$0.11 / $\leq$0.068 &  4(-10) & $\leq$ 5(-9)&$\leq$3(-8)\\
6933 & 50 &  38100  & 10 &   0.11 &  0.100 &  $\leq$0.11 / $\leq$0.068 &  5(-10) &  $\leq$5(-9)&$\leq$3(-8)\\
8666 & 62 &  33900  & 10 &   0.11 &  0.089 &  $\leq$0.11 / $\leq$0.068 &  5(-10) &  $\leq$5(-9)&$\leq$3(-8)\\
10400 & 75 &  24300  & 10 &   0.11 &  0.082 &  $\leq$0.11 / $\leq$0.068 &  4(-10) & $\leq$ 5(-9)&$\leq$3(-8)\\
12133 & 87 &  18500  & 10 &   0.11 &  0.076 &  $\leq$0.11 / $\leq$0.068 &  3(-10) &  1(-8)&6(-8)\\
13933 & 100 &  14700  & 10 &   0.11 &  0.070 &  $\leq$0.11 / $\leq$0.068 &  2(-10) &  1.7(-7)&1(-6)\\
15666 & 112 &  11900  & 9 &   0.11 &  0.066 &  $\leq$0.11 / $\leq$0.068 &  1(-10) &  1.7(-7)&1(-6)\\
17400 & 125 &  9990  & 8 &   0.11 &  0.063 &  $\leq$0.11 / $\leq$0.068 &  1(-10) &  1.7(-7)&1(-6)\\
19133 & 137 &  8540  & 8 &   0.11 &  0.060 &  $\leq$0.11 / $\leq$0.068 &   &  1.7(-7)&1(-6)\\
20866 & 150 &  7400  & 8 &   0.11 &  0.058 &  $\leq$0.11 / $\leq$0.068 &   &  1.7(-7)&1(-6)\\
22600 & 162 &  6100  & 8 &   0.11 &  0.055 &  $\leq$0.11 / $\leq$0.068 &   &  1.7(-7)&1(-6)\\
24333 & 175 &  4800  & 8 &   0.11 &  0.053 &  $\leq$0.11 / $\leq$0.068 &   &  1.7(-7)&1(-6)\\
26066 & 187 &  3850  & 8 &   0.11 &  0.052 &  $\leq$0.11 / $\leq$0.068 &   &  1.7(-7)&1(-6)\\
27866 & 200 &  3130  & 8 &   0.11 &  0.050 &  $\leq$0.11 / $\leq$0.068 &   &  1.7(-7)&1(-6)\\
29600 & 212 &  2560  & 8 &   0.11 &  0.048 &  $\leq$0.11 / $\leq$0.068 &   &  1.7(-7)&1(-6)\\
31333 & 225 &  2140  & 8 &   0 &  -0.3 & 0.4 &   &  $\leq$5(-9)&6(-6)\\
46666 & 335 &  1000  & 8 &   -0.09 &  -0.3 & 0.4 &   &  $\leq$5(-9)&6(-6)\\
66666 & 479 &  500  & 10 &   -0.09 &  -0.3 &  0.4 &   &  $\leq$5(-9)&6(-6)\\
83333 & 598 &  400  & 12 &  -0.09 &  -0.3 &  0.4 &   &  $\leq$5(-9)&6(-6)\\
100000 & 718 &  400  & 13 &   -0.09 &  -0.3 &   0.4  &   & $\leq$5(-9)&6(-6)\\
126666 & 909 &  200  & 13 &   -0.09 &  -0.3 &   0.4  &   &  $\leq$1(-9)&1(-8)\\
\hline
\end{tabular}
\begin{list}{}{}
\item[$^{\mathrm{a}}$]positive velocity indicates inward motion (contraction). 
\item[$^{\mathrm{b}}$] FWHM turbulence is $\leq$0.068\kmpsb for the positive offsets in the \cdhob model and for all offsets in the \ndhpb model {and $\leq$ 0.11 \kmpsb otherwise} (for Figs. \ref{fig:fitc18o10} and  \ref{fig:fitc18o21} only one value could be used: 0.11 \kmpsb has been selected).
\item[$^{\mathrm{c}}$] relative to H$_2$. 1(-10) means 1 $\times$ 10$^{-10}$.
\end{list}
\end{table*}%

The CO+\ndhpb model which we present here is by no means unique but represents a family of possible solutions. They have remarkable features which are common to all of them:
\begin{enumerate}
\item a peak column density of N(H$_2$) $\approx$ 2 \pdix{22} \cde, equivalent to \Av $\approx$ 20 mag \citep[following][]{1978ApJ...224..132B}. This is low compared to most other cases with similar strong \cdhob depletion but slightly higher than in L1498 which has a peak column density of N(H$_2$) $\approx$ 1.3 \pdix{22} \cde\ \citep{2005ApJ...632..982S}\footnote{Note that \citet{2005ApJ...632..982S} disagree on the total dust amount and peak density of L1498 with respect to what \citet{2004A&A...416..191T} find. If we follow the latter, L1498 is closer to a normal prestellar core than to a simple core. The discrepancy has only partly been explained by \citet{2005ApJ...632..982S}.}. 
\item a strong central depletion ($\geqslant$ 30) of \cdhob.
\item \cdhob depletion starting at a density between 1.5 and 2 \pdix{4} \cc.
\item for both the \cdhob lines (towards the positive offsets) and the \ndhpb line (for all offsets), the linewidth is so narrow that the turbulence contribution to the width is negligible and is set to V$_\mathrm{turb}\mathrm{(FWHM)} \leqslant$ 68 m\,s$^{-1}$. This is the lowest turbulence reported for \cdhob lines to our knowledge. The line width is mostly dominated by the thermal width and the enlargement due to macroscopic kinematics (rotation and infall).
\item the radial velocity is strongly constrained in the \cdhob envelope but not so much in the \ndhpb internal part. However, a constant infall around 0.1 \kmpsb is compatible with the data. No infall is not.
\item {the rotational velocity is decreasing with increasing radius. Solid rotation is therefore excluded and velocity dependences of the type V $\propto r^{-\alpha}$ with $\alpha \approx 0.5$ gave good solutions}.
\item the temperature pattern requests a slightly higher temperature in the \ndhpb inner core ($\sim$10 K) than in the \cdhob outer core ($\sim$8 K). If correct this would be similar to the B68 case \citep{2006ApJ...645..369B}.
\end{enumerate}

{The \cdhob temperature is determined by the necessity to fit both the \juzb and \jdub lines with a relatively constrained dust profile while the \ndhp temperature is the minimum we could get trying to minimize also the needed density. Despite the observation of a single transition for \ndhp, we are constrained by the hyperfine structure \citep[see ]{2007A&A...467..179P}. However this temperature difference }could possibly prove to be inaccurate if the actual H$_2$--\ndhpb collisional coefficients were available: they are expected to be higher than the present He--\ndhpb ones and therefore would probably allow to fit the observations with both a lower temperature and a lower density. In any case, the gas appears cooler than what \citet{2003A&A...398..551S} have found for the dust ($\sim12$\,K). Either their resolution is too coarse and mix up a large fraction of the (warmer) envelope emission with the core emission, or their single temperature fit is too simple by lack of constraints \citep[we have shown that in the case of \object{L183} cold core the 200 $\mu$m emission is not tracing the same dust as the submm emission and that a temperature gradient is compulsory,][]{2004A&A...417..605P} or, finally the gas and the dust being decoupled at these low densities (n(H$_2$) $\le$ 5 \pdix{4} \cc\ or less), the dust could be warmer than the gas though this is not usually expected.
 
The modeling of the \nddpb \juzb line indicates a column density of 4 to 8 \pdix{11} \cde, depending on the chosen \nddpb abundance profile which we cannot constrain with only one observation. The peak \ndhpb \cdb is $\sim$2 \pdix{13} \cde\ and therefore the deuterium enrichment is 2 -- 4 \%. Compared to L183 \citep{2007A&A...467..179P,2008arXiv0810.1861P}, we find a similar deuteration level at similar densities indicating that the evolutionary speed of the deuterium enrichment is probably similar and therefore can be used to determine the age of the core.

Beyond the core characterized by \cdhob narrow lines, the envelope, as traced by \tzco, is best described with an almost constant density (in the range 200--1000 \cc) and a large turbulence (V$_\mathrm{turb}\mathrm{(FWHM)}$  = 0.4 \kmps). To follow its macroscopic velocity drift we had to introduce a counter-rotating pattern in the external layers set constant to -0.3 \kmps. We do not claim that the \tzcob is indeed rotating but this is the only way to approach the velocity profile of this line with our model. Similarly, to reproduce the redshifted peak, we had to introduce an expansion velocity of 90 m\,s$^{-1}$, that is comparable in amplitude to the infall velocity of the core. The fit is not perfect inasmuch as we could not centre the peak emission in our model at the same position as in the observed position-velocity diagram (Fig. \ref{fig:VPobsmod}) but the fit bears some ressemblance with the reality as shown also in Fig. \ref{fig:fit13co10}. The derived parameters are therefore reasonably close to the actual situation and this envelope is clearly decoupled from the core in terms of density profile as much as in terms of turbulence and large scale motions. That region is also marked by a very high abundance of \tzcob (6 \pdix{-6}) and the absence of \cdhob ($<$ 5 \pdix{-9}, an abundance of 1 \pdix{-8} is marginally consistant with the observations). The density is not well constrained in the external envelope but as the \tzcob abundance is already high (standard value is 1--2 \pdix{-6}, i.e. 3 to 6 times less), the density cannot be lower than what we have assumed here as a lower density would increase even more the \tzcob abundance. Conversely, we could increase the density to lower the abundance of \tzco, but it would become higher than the density in the last  \cdhob layers. In the absence of other \tzcob transitions it is difficult to constrain correctly the density and the abundance in these last, turbulent layers.

The collapsing core has a size of $\sim$3 \pdix{4} AU in radius and holds a mass of $\sim$4 M$_{\sun}$. This size is larger than the one of L1498 which was already shown to be the largest by 50\% among the nearby prestellar cores mentioned in the study by \citet{2005ApJ...632..982S}. Our mass estimate is somewhat approximate as the density of the \ndhpb region is possibly overestimated but on the other hand, the core could be slightly more extended than a sphere, closer to a cylindar. The estimate derived from the dust observations is similar but less constrained due to the unknowns on the dust emissivity and temperature in the core at scales smaller than the PRONAOS resolution. The \tzcob low density filament holds at least 20 M$_{\sun}$ accounting only for the part along its main axis which covers the core. Its diameter is 250,000 A.U. or 1.25 pc.

\section{Discussion}

\begin{figure}[htbp]
   \flushleft
\includegraphics[width=6.8cm,angle=-90]{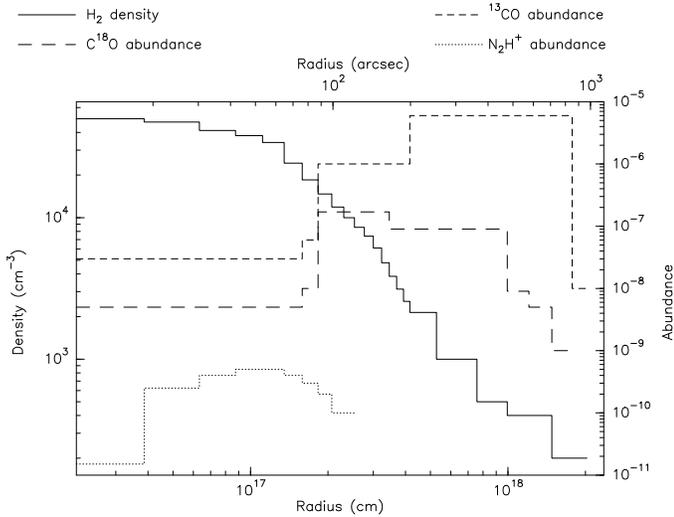}
\caption{H$_2$ density and molecular abundance (relative to H$_2$) as a function of radius.}
\label{fig:profil}
\end{figure}

Several aspects of this part of the L1506 filament are surprising and presently difficult to explain. The expanding and counter-drifting external envelope is one of them. One partially similar case has however already been reported: \citet{1981ApJ...251L..81Y} have shown that the \object{B5} core as traced by \cdhob had a position/velocity profile gradient opposite to the \dzcob one which they interpreted as a counter-rotating core (but they don't discuss the radial component). Contemporary theoretical work \citep{1979MNRAS.187..311G,1979ApJ...230..204M} was invoked to explain this situation: a radial, frozen-in magnetic field perpendicular to the rotation axis could brake down the rotation of the cloud but because the field remains anchored to the external regions, the magnetic torques do not vanish immediately and counter-rotation is induced in the core which could oscillate several times before coming to a halt. Whether this is a similar case here is not yet clear because 1) we are not certain that the \tzcob itself is tracing a rotation, 2) the expansion of the \tzcob envelope is not predicted by the model and, 3) we know nothing about the magnetic field inside this core \citep[external parts have been mapped though,][]{1984ApJ...282..508M,2000AJ....119..923H}.

The strong depletion of \cdhob with its abrupt appearance at relatively low density is another one. It is similar to the L1498 case but in sheer contrast with other very young cores, like L1495B, L1521B, and L1521E which show no depletion despite a density which is typically 5 times higher \citep[and possibly 10 times if He-\ndhpb collisional coefficients are indeed over-evaluating the actual density in L1506C or if we compare to L1498 peak density as derived by][]{2005ApJ...632..982S}. Based mainly on this absence of depletion, these cores have been judged to be young but it is somewhat strange that having very recently acquired their present density state, they show no kinematical sign of contraction. It is thus conceivable that they are older than assumed and that the absence of depletion could be due to another mechanism, like a strong desorption, an ill-understood mechanism \citep{2007MNRAS.382..733R}. Indeed, it is thought that energetic cosmic ray impacts can desorb the most volatile ices like CO \citep{1985A&A...144..147L,1993MNRAS.261...83H, 2004ApJ...603..159B,2007MNRAS.382..733R}. The desorption efficiency is inversely correlated to the size of the grains as smaller grains are easier to warm up. Therefore the question arises whether cores like L1495B have smaller grains than L1506C or not and how much different in size should these grains be to allow or prevent an efficient desorption of CO. As explained by \citet{2007MNRAS.382..733R}, all the desorption mechanisms are yet poorly known and it is presently difficult to go further. However, if this explanation proves to be valid, then one has to understand why grains coagulate in some sources and not in others. 
{Turbulence could be one of the main key parameters involved in this differentiation between cores. From \citet{1993ApJ...407..806C} and \citet{1997AdSpR..20.1595P} works, \citet{2005A&A...436..933F} show that dust coagulation efficiency peaks for turbulent velocities in the range 10 -- 80 m\,s$^{-1}$ (see their Fig. A.1), a condition that is hardly met even in the most quiescent prestellar cores but fits the present case. On the opposite, L1521E shows no depletion while the turbulence is suprathermal \citep[even for \ndhp, V$_\mathrm{turb}\mathrm{(FWHM)}$  = 0.27 \kmps,][ a factor of 4 at least larger than in L1506C]{2004A&A...414L..53T}. If the size of the grains in L1521E is not known, large grains have been advocated to explain sub-millimeter emission in L1506C:  \citet{2003A&A...398..551S} have built a model where the very small grains, emitting at 60 $\mu$m, disappear for offsets lower than $r_0=4$\arcmin\ from the centre while the sub-mm emissivity increases by a factor of 3.4 due to the formation of fluffy aggregates. Their observations were taken with an angular resolution of 2-3.5', so the value of $r_0$ could be lower and fit the extent ($\pm$ 200\arcsec) of the low turbulence region in L1506C. }
Grain growth, depletion and desorption are certainly not completely understood at the moment.

The absence of \cdhob on 5/6th of the external envelope width as traced by the POM-1 \tzcob observations is another puzzle. The \tzcob abundance itself seems to be higher than usual so that the \tzco/\cdhob ratio reaches 600--1200 over a large region. Though \tzcob fractionation does exist, it is weak and does not seem to be able to explain this difference. More observations of the envelope (including \tzcob \juzb and \jdub and very low noise \cdhob transitions) are needed to address this issue.

Clear trace of collapse among prestellar cores has been found in a few cases. \citet{1999ApJ...526..788L, 2004ApJS..153..523L} observed a large number of asymmetrically blue profiles towards a sample of starless cores but they traced the CS kinematics with respect to \ndhpb or DCO$^+$ and, as CS is depleted inside the \ndhpb region, they trace the inward motion of the envelope only. A few prestellar cores are indeed modeled with a collapsing core, the best studied one being \object{L1544} \citep[e.g.][]{1999ApJ...513L..61W,2002ApJ...569..815T, 2005A&A...439..195V} which was reported to clearly undergo a collapse from line profile fitting.  \citet{2004A&A...416..191T} also invoke the possibility of infall from the observations of two other cores, \object{L1498} and \object{L1517B} despite the absence of a clear velocity gradient and  \citet{2006ApJ...636..952W} found that \object{L694-2} has a profile reminiscent of L1544, indicating also probable infall. This has been confirmed by \citet{2007ApJ...660.1326L} who also report infall of \object{L1197}. Here the inward motion of \ndhpb is not strongly constrained in amplitude but unavoidable to reproduce the velocity-profile plot{: no infall at all gives clearly a wrong result but the details of the radial velocity profile are uncertain}. The inward motion is also clearly shown by the \cdhob and \cdsob lines. It is the first time to our knowledge that such lines do trace the collapse of a core . We think that this is possible here because several favorable factors are met :
\begin{enumerate}

\item no CO isotopologues in the central part, separating clearly the front and rear moving layers as indicated by \cdsob.
\item no \cdhob in the extended envelope which would have hidden or blurred the faint variation of intensity between the blue and red emissions which let us differentiate between expansion and contraction
\item low \cdhob velocity dispersion, increasing the line opacity near unity ($\tau_{J:1-0}$ = 0.56, $\tau_{J:2-1}$ = 0.91). Combined with the outward decreasing excitation temperature, it allows for the weak but measurable differential absorption between the blue and red components.
 
\end{enumerate}
The most surprising aspect is that this core with a density below $\sim$1 \pdix{5}~\ccb is supposed to be dynamically stable given its total mass \citep[less than 10 \SM,][]{2008ApJ...683..238K} which is clearly not the case here (the core mass being estimated to be $\sim$4 \SM). However, as noted by \citet{2008ApJ...683..238K}, most clouds without non-thermal energy support would be unstable and in this case, turbulence is abnormally low, non-measurable in several parts of the core. This would support the idea of a possible collapse though we have no clue on the magnetic field intensity and direction. If a magnetic field is indeed present to induce the core counter-rotation, it is not incompatible with  infall as it is included in the magnetic braking model of \citet{1979ApJ...230..204M}. Oscillations or sound waves have been invoked in the case of e.g. B68 \citep{2007ApJ...670L..25M} and one could be tempted to invoke a similar argument here but the rotation of \cdhob and "counter-rotation" of \tzcob seem difficult to explain by sound waves.

\section{Conclusions}

The large size of L1506C and its low density clearly indicate that it is not yet a prestellar core but its inward motion and its kinematical decoupling with the outer envelope are signs that it is in the process of turning into one. It is the first time that a low density core is clearly observed to collapse to form a prestellar core, both the kinematical and density status of L1498 being somewhat unclear. Two different evolutionary paths from normal clouds to prestellar cores seem to exist : one which starts with density enhancement first much before any depletion occurs and one which starts with depletion first. Two questions arise : what is the reason for this differentiation {(turbulence via its action on grain growth, itself limiting the desorption capability of cosmic rays ?)} ? and are both paths really leading to standard prestellar cores ? Coming observations with the Herschel Space Observatory will help to address these questions and the nature of the dust inside this object.

\begin{acknowledgements}
We want to thank an anonymous referee for her/his fruitful comments which helped to improve the manuscript and P.F. Goldsmith for fruitful discussions.
\end{acknowledgements}
\bibliographystyle{aa}
\bibliography{/Users/laurent/Bibtex/References}
%:bibliographie
%\begin{thebibliography}{}
%\end{thebibliography}

\end{document}